\documentclass[useAMS,usenatbib]{mnras}

\usepackage{amsmath} 
\usepackage{amssymb} 
\usepackage{graphicx}

\title[Troughs in the Ly$\alpha$ forest and \ion{H}{i} islands]{Long troughs in the Lyman-$\alpha$ forest below redshift 6 due to islands of neutral hydrogen}

\author[L.C. Keating et al.]{\parbox{\textwidth}{Laura C. Keating$^{1}$\thanks{E-mail: lkeating@cita.utoronto.ca}, Lewis H. Weinberger$^{2,3}$, Girish Kulkarni$^{4}$,\\ Martin G. Haehnelt$^{2,3}$, Jonathan Chardin$^5$ and Dominique Aubert$^5$}\vspace{0.4cm} \\
\parbox{\textwidth}{$^1$Canadian Institute for Theoretical Astrophysics, 60 St. George Street, University of Toronto, ON M5S 3H8, Canada\\
$^2$Institute of Astronomy, University of Cambridge, Madingley Road, Cambridge CB3 0HA, UK\\
$^3$Kavli Institute for Cosmology, University of Cambridge, Madingley Road, Cambridge CB3 0HA, UK\\
$^4$Department of Theoretical Physics, Tata Institute of Fundamental Research, Homi Bhabha Road, Mumbai 400005, India\\
$^5$Observatoire Astronomique de Strasbourg, 11 rue de l'Universite, 67000 Strasbourg, France\\}
}

\begin{document}

\date{\today}

\pagerange{\pageref{firstpage}--\pageref{lastpage}} 

\pubyear{2019}

\maketitle

\label{firstpage}

\begin{abstract}
A long (110 cMpc/$h$) and deep absorption trough in the Ly$\alpha$ forest has been observed extending down to redshift 5.5 in the spectrum of ULAS J0148+0600. Although no Ly$\alpha$ transmission is detected, Ly$\beta$ spikes are present which has led to claims that the gas along this trough must be ionized. Using high resolution cosmological radiative transfer simulations in large volumes, we show that in a scenario where reionization ends late ($z \sim 5.2$), our simulations can reproduce troughs as long as observed. In this model, we find that the troughs are caused by islands of neutral hydrogen. Small ionized holes within the neutral islands allow for the transmission of Ly$\beta$. We have also modelled the Ly$\alpha$ emitter population around the simulated troughs, and show that there is a deficit of Ly$\alpha$ emitters close to the trough as is observed.
\end{abstract}


\section{Introduction}

The Lyman-$\alpha$ (Ly$\alpha$) forest has long been recognised as a useful probe of the tail end of the epoch of reionization. Gunn-Peterson troughs observed towards redshift 6 are interpreted as a sign the intergalactic medium (IGM) is becoming increasingly neutral towards these redshifts. However, the Ly$\alpha$ forest saturates for relatively low \ion{H}{i} fractions meaning that other methods must be used to place constraints on reionization. One such observation is the large scale spatial fluctuations in the opacity of the Ly$\alpha$ forest. These were first noted by \cite{fan2006}, however \citet{lidz2006} claimed that the fluctuations could be explained by fluctuations in the density field alone. More recently, larger sets of higher quality observations that push to higher redshifts have been obtained \citep{becker2015,bosman2018,eilers2018}. These observations show even larger fluctuations in the opacity of the Ly$\alpha$ forest at a given redshift. Furthermore, \citet{becker2015} found a 110 cMpc/$h$ trough in the Ly$\alpha$ forest that showed complete absorption in Ly$\alpha$, although Ly$\beta$ transmission was observed. These observations could not be explained by variations in the density field and were claimed to be a signature of patchy hydrogen reionization.

Several competing models were proposed to explain how the \ion{H}{i} fraction could vary over such large scales. \citet{chardin2015, chardin2017} proposed a contribution to the UV background (UVB) from rare, bright sources such as AGN. \citet{davies2016} suggested that the mean free path of ionizing photons may be very short in underdense regions, allowing for large scale fluctuations in the mean free path which in turn drive fluctuations in the amplitude of the UVB. \citet{daloisio2015} suggested that the temperature dependence of the recombination rate may allow for a difference in \ion{H}{i} fractions between regions that ionized early and have had time to cool compared with regions that ionized late and are still hot. 

Although each of these models can potentially explain the distribution of Ly$\alpha$ forest opacities, they each require some key assumption which has been challenged. The model requiring rare, bright sources could lead to an early \ion{He}{ii} reionization inconsistent with observations \citep[e.g.,][]{daloisio2017,garaldi2019}. The fluctuating UVB model requires a mean free path at $z > 5$ much shorter than observed at slightly lower redshift \citep{worseck2014}. For the temperature fluctuations model, gas is required to reach very high temperatures as it is ionized which appear to be inconsistent with measurements of the IGM temperature above redshift 5 \citep[e.g.,][]{bolton2012,walther2019}. 

There is however a way to discriminate between these models, as pointed out by \citet{davies2018}. In the case where the fluctuations are primarily in the gas temperature, the most overdense regions of the IGM will have the highest opacity as they were ionized first. In the case where the fluctuations in the amplitude of the UV background dominate, the high opacity regions will instead be the low density regions that are far from sources. This was tested by \citet{becker2018}, who used a narrowband filter on Subaru Hyper-Suprime Cam (HSC) to search for Ly$\alpha$ emitters (LAEs) in the vicinity of the long trough of ULAS J0148+0600. They found an underdensity of LAEs around the trough, suggesting that the sightline lay in a low density environment. This is consistent with the \citet{davies2016} model, where a strong density dependence of the mean free path results in a rather low amplitude of the UVB in low density regions.

\citet{kulkarni2019} recently showed that another way to explain the fluctuations in the opacity of the Ly$\alpha$ forest is if reionization ended later than previously thought \citep[see also][]{lidz2007,mesinger2010}. Using radiative transfer simulations of the reionization of the IGM, they found a reionization history that was consistent with Ly$\alpha$ forest constraints on the ionization state of the IGM below redshift 6 \citep{mcgreer2015,bosman2018,eilers2018} as well as constraints from the low optical depth to Thomson scattering measured with the CMB \citep{planck2018}. In this model, there were contributions from fluctuations in the UVB before reionization ends, as well as from temperature fluctuations in the IGM. However to reproduce the sightlines with the highest effective optical depths, it was crucial that residual islands of neutral hydrogen persisted down below redshift $z \leq 5.5$. 

In this work we focus on explaining how the long trough of ULAS J0148+0600 can be explained in this late reionization model. In section 2, we present results from high resolution radiative transfer simulations that were tuned to match the mean flux of the Ly$\alpha$ forest below redshift 6. In section 3, we show that, as well as matching the distribution of mean fluxes measured along individual sightlines, these models can reproduce troughs as long as observed. In section 4, we model the distribution of LAEs around one of our simulated troughs and in section 5 we present a test to distinguish between a late reionization model and a scenario with a fluctuating UVB. Finally in section 6 we state our conclusions.

\section{Modelling reionization}

\begin{figure}
\includegraphics[width=\columnwidth]{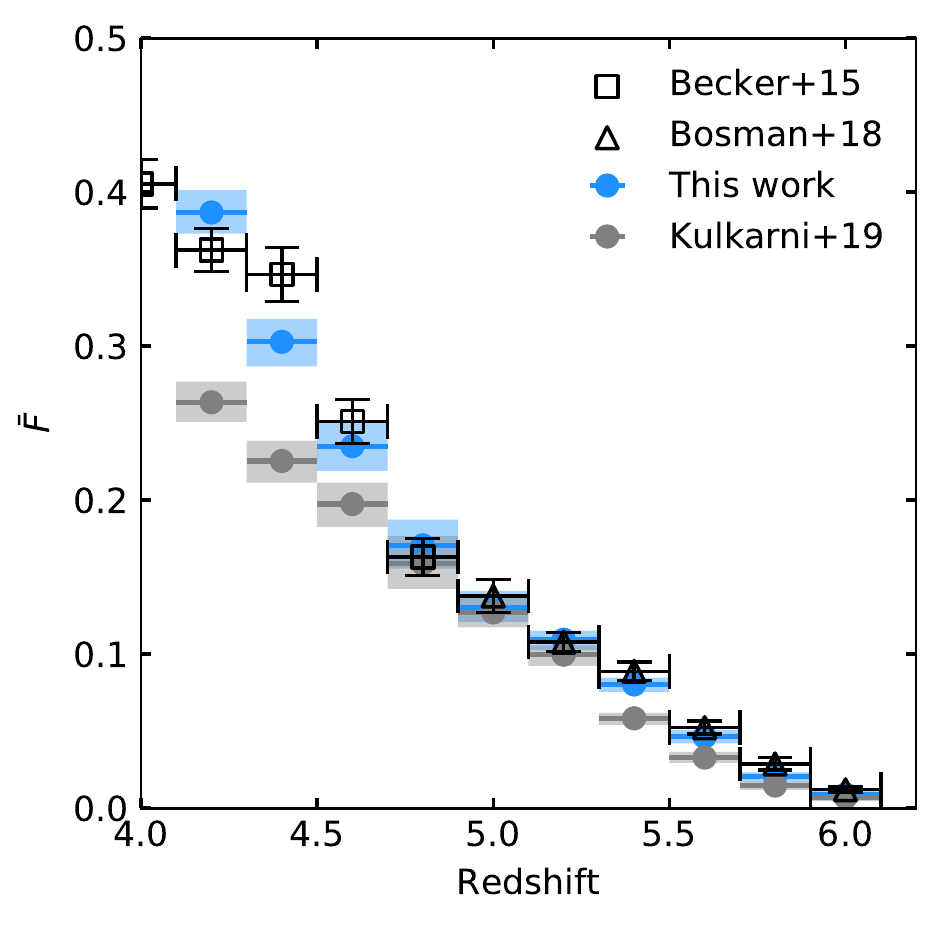}
\caption{The evolution of the mean flux (blue points) in our simulation down to $z \sim 4$ compared to observations from \citet{bosman2018} at high redshift and \citet{becker2015} at lower redshift. For comparison, we also show the mean flux of the simulation presented in \citet{kulkarni2019} (grey points) measured in the same way as in this work. As we construct these spectra from using sightlines along lightcones through the box, rather than by interpolating between snapshots at different redshifts, there is a small difference between the mean flux quoted here for \citet{kulkarni2019} and what was presented in that work.}
\label{meanflux}
\end{figure}

To model the properties of the IGM at the end of reionization, we present here a carefully calibrated simulation of hydrogen reionization. We ran mono-frequency cosmological radiative transfer simulations in post-processing using \textsc{aton} \citep{aubert2008,aubert2010}, using a setup similar to \citet{chardin2015}. \textsc{aton} is a GPU-based code that solves the radiative transfer equation using a moment-based method with M1 closure for the Eddington tensor \citep[e.g.,][]{levermore1984,gnedin2001,aubert2008}. We perform this post-processing on a set of density fields obtained from a hydrodynamic simulation run with Tree-PM SPH code \textsc{p-gadget3} \citep[last described by][]{springel2005gadget}. Our simulation has a box size 160 cMpc/$h$ and 2$\times$2048$^{3}$ gas and dark matter particles. The initial conditions were taken from the Sherwood simulation suite \citep{bolton2017} and assume cosmological parameters from \citet{planck2014} with $\Omega_{\rm m} = 0.308$, $\Omega_{\Lambda} = 0.692$, $h = 0.678$, $\Omega_{\rm b} = 0.482$, $\sigma_{8} = 0.829$ and $n = 0.961$. The gas particles have mass $m_{\rm gas} = 6.4 \times 10^{6}$ $M_{\odot}$/$h$ and the dark matter particles have $m_{\rm DM} = 3.4 \times 10^{7}$ $M_{\odot}$/$h$. The gravitational softening length we assume is 3.1 ckpc/$h$. We run these simulations using a simplified star formation scheme, whereby any gas particles with a density greater than 1,000 times the mean cosmic baryon density and temperature less than 10$^{5}$ K are immediately converted into stars \citep{viel2004}. This allows us to lower the time required to run the simulation without significantly affecting the properties of the intergalactic medium. 

As we are performing the radiative transfer in post-processing, we neglect the hydrodynamical back-reaction of the gas to the photoheating. To compensate for this by adding some pressure smoothing to our input density fields, we also include photoheating by the \citet{haardtmadau2012} uniform UV background in our hydrodynamic simulation. We start the simulations at redshift 99 and run them down to redshift 4. To allow us to update our input density field in the radiative transfer simulations frequently, we take snapshots of the simulation every 40 Myr starting just below redshift 20. We search for haloes in the simulation on the fly using a friends-of-friends algorithm. As \textsc{aton} requires the input density fields to be described on a cartesian grid, we map the SPH particles onto a 2048$^{3}$ grid (with cell size 78.125 kpc/$h$) using the SPH kernel. To assign sources in the simulation, we assume that each halo has an emissivity proportional to its mass \citep{iliev2006b}. The emissivity of each source is determined by summing the total mass of sources within our volume and dividing the assumed total emissivity within our volume among them, proportional to their masses. We take a minimum halo mass $10^9$ $M_{\odot}$/$h$. We find changing this minimum mass has hardly any effect on our results (once we have recalibrated the simulations to match the mean flux of the Ly$\alpha$ forest).

\begin{figure*}
\includegraphics[width=2\columnwidth]{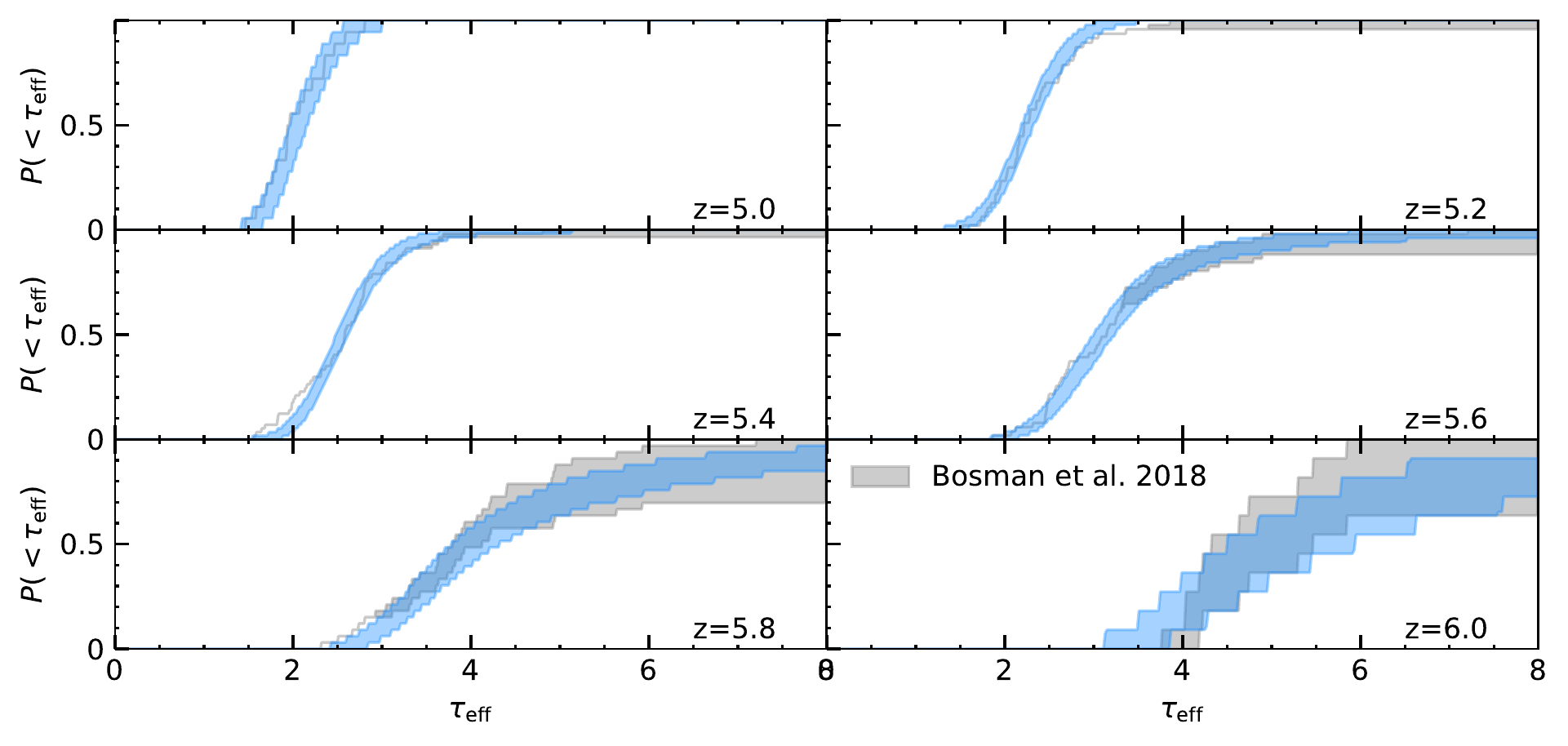}
\caption{The cumulative distribution of effective optical depths in our model (blue region), compared to the observations from \citet{bosman2018} (grey region). For the model, the shaded region denotes the 15$^{\rm th}$ and 85$^{\rm th}$ percentiles of 500 realisations of the CDFs. For the observations, the shaded area represents the ``optimistic'' and ``pessimistic'' bounds for the data.}
\label{opacities}
\end{figure*}

We calibrate the ionizing emissivity assumed in the simulation by running many simulations until the properties of our simulated Ly$\alpha$ forest (described later) agree well with the observations. As described above, the emissivity of each source is completely determined by the mass of the host halo and the volume emissivity we assume. When we calibrate the simulations, we only change the volume emissivity. This is equivalent to assuming some evolution in the properties of the sources, such as their escape fraction, but we do not explicitly treat these parameters.  We make some changes with respect to the simulation presented in \citet{kulkarni2019}, namely using a lower photon energy and a somewhat different emissivity evolution. The starting point for the emissivity evolution, which we modify as needed, is a galaxy-like component from \citet{puchwein2019} and an increasing AGN-like contribution at lower redshifts. Although we do not model ionization by AGN directly (we only model one class of source, with a spectrum assumed to be galaxy-like), we find that we need an increasing volume emissivity at lower redshifts to match the mean flux of the Ly$\alpha$ forest (described further below). We do not try to separate our volume emissivity into contributions from galaxy-like and AGN-like components, but only account for the total emissivity that we require to reproduce the Ly$\alpha$ forest statistics. We therefore leave a more detailed comparison of the relative contributions of galaxies and AGN to the high-redshift UVB to future work. During reionization, we assume that all of our sources emit as 30,000 K blackbodies. This spectrum was chosen to provide reasonable agreement with measurements of the IGM temperature below redshift 6. This is equivalent to an average photon energy of 17.1 eV and a photoionization cross-section $3.9 \times 10^{-18}$ cm$^{-2}$. We arrive at these numbers assuming the optically thick limit, assuming all photons are absorbed locally \citep{pawlik2011,keating2018}. This is likely a good approximation during reionization but will cause us to overestimate our photoheating after reionization has ended. For this spectrum, gas that is instantaneously ionized would reach a maximum temperature of approximately 13,000 K.

\begin{figure*}
\includegraphics[width=2\columnwidth]{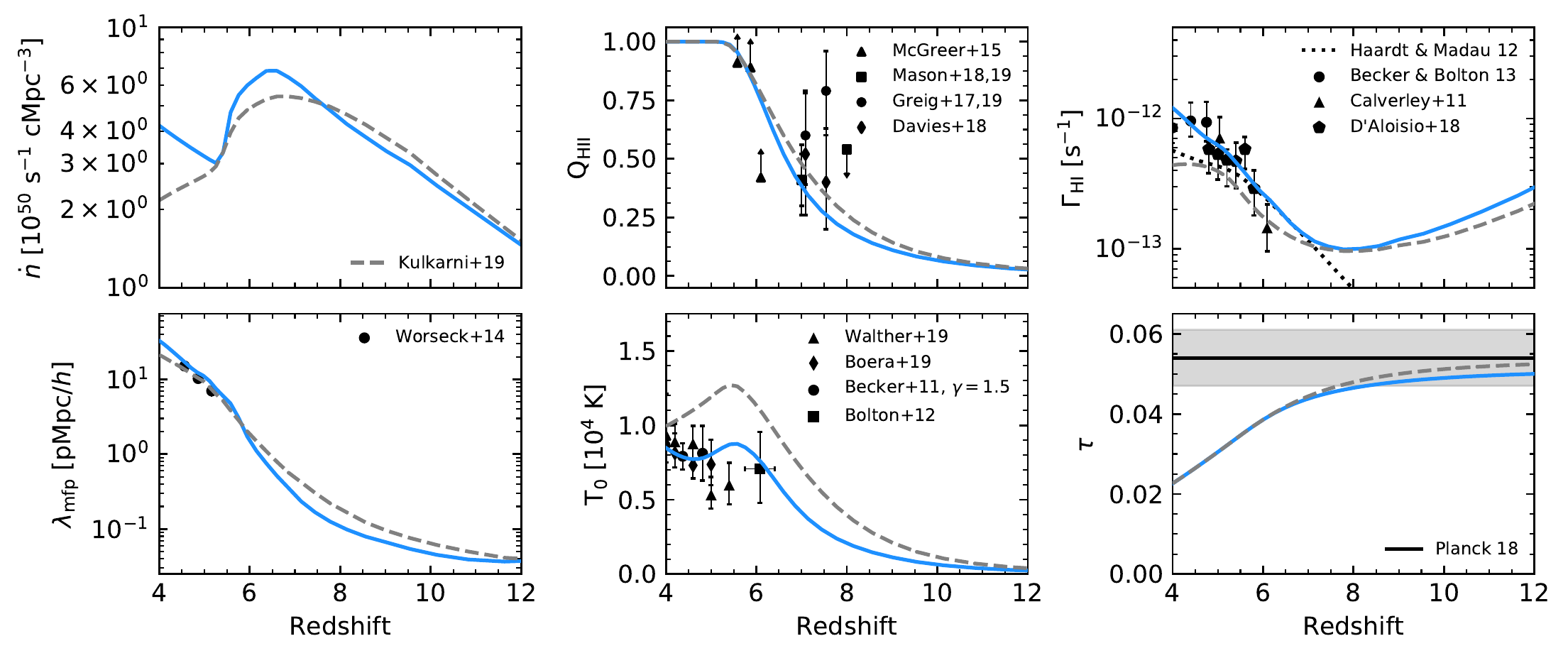}
\caption{Properties of our radiative transfer simulation (blue line) compared with the simulation presented in \citet{kulkarni2019} (grey dashed line). Top left: The evolution of the ionizing emissivity with redshift. Top middle: The filling fraction of ionized gas. Also shown are estimates from dark gap statistics of the Ly$\alpha$ forest \citep{mcgreer2015}, quasar damping wing studies \citep{greig2017,greig2019,davies2018} and estimates from the fraction of Lyman break galaxies that show Ly$\alpha$ emission \citep{mason2018,mason2019}. Top right: The evolution of the \ion{H}{i} photoionization rate measured in ionized regions compared with measurements from \citet{becker2013}, \citet{calverley2011} and \citet{daloisio2018}. The black dotted line is the \citet{haardtmadau2012} model. Bottom left: The evolution of the mean free path for photons at 912 \AA \, as well as measurements from \citet{worseck2014}. Bottom middle: The mean temperature at mean density compared with estimates from \citet{walther2019}, \citet{boera2019}, \citet{becker2011temp} and \citet{bolton2012}. Bottom right: The Thomson optical depth compared with the latest \textit{Planck} results \citep{planck2018}.}
\label{history}
\end{figure*}

At lower redshifts, in a crude attempt to mimic the effects of the heating of the IGM by \ion{He}{ii} reionization \citep[e.g.,][]{becker2011temp}, we increase the photon energy in our simulation. We do not explicitly treat two classes of sources (galaxies and AGN) separately, but instead just try to capture the average effect of the evolution of the shape and amplitude of the UVB on the IGM. This change in the photon energy therefore applies to every photon in the simulation (both those emitted from sources and from recombinations). We begin increasing the energy at redshift 5.2 linearly, and by redshift 4 we use an average photon energy of 23.8 eV. For the intermediate redshifts we interpolate between the values at $z=5.2$ and 4 linearly with the cosmic expansion scale factor. Increasing the photon energy in this way is fully capturing the relevant physics, as any heating of the gas at these redshifts will come from recombinations, and we are not properly modelling e.g. the extra heating of the IGM by X-rays, so it is likely we will not recover an appropriate temperature-density relation as our voids will be under-heated. We also change the photon energy of photons that have already been emitted from sources, which will not conserve energy. We also neglect the corresponding change in the photoionization cross-section as we harden our spectra, as we wish to heat the gas without changing the photoionizing background.  However, we found it was hard to recover the observed mean flux at redshift 4 if our gas was too cold and so decided to adopt this approximate treatment despite the shortcomings discussed above.  We use the full speed of light in our simulations and take a timestep $\Delta t < \Delta x / c$  where $c$ is the speed of light and $\Delta x$ is the cell size of the grid. The chemical network only follows the photoionization, photoheating and cooling of hydrogen. The cooling processes we include are Hubble cooling, collisional ionization and excitation cooling \citep{cen1992}, recombination cooling \citep{huignedin1997}, Compton cooling \citep{haiman1996} and free-free cooling \citep{osterbrock2006}.

As we run the simulation, we extract the ionization state and temperature of the gas on the fly along a lightcone taken at a 30 degree angle through our grid. This allows us to account for the rapid evolution of the IGM along a line of sight, rather than having to interpolate between individual snapshots of the simulation as in \citet{kulkarni2019}. We then construct mock Ly$\alpha$ forest spectra along sightlines along our output lightcone. The optical depth was computed using the approximate fit to Voigt profiles presented in \citet{teppergarcia2006}. We show the evolution of the mean flux in our simulations in Figure \ref{meanflux} compared with observations from \citet{becker2015} and \citet{bosman2018}. In this work, we only focus on analysis of the Ly$\alpha$ forest down to $z \sim 5.5$ (the end-point of the \citet{becker2015} long trough). However, as we also wish to look at the Ly$\beta$ transmission along our simulated troughs, it is important to get the mean flux of the Ly$\alpha$ forest close to the observed values at lower redshifts for adding in the contamination from the foreground Ly$\alpha$ forest.

\begin{figure*}
 \includegraphics[width=2\columnwidth]{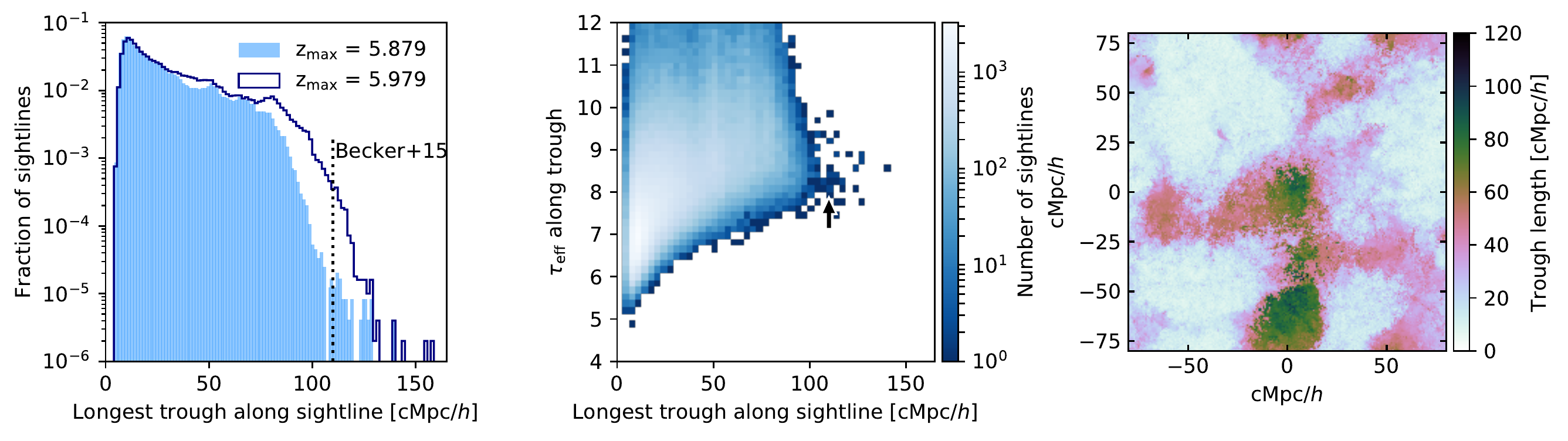}
\caption{Left panel: Distribution of the longest trough we measure along a sample of sightlines through our simulation. The filled blue histogram only shows troughs with a maximum redshift $z \leq 5.879$, the high redshift endpoint of the \citet{becker2015} trough. The empty navy histogram shows another search for troughs where we have increased the maximum redshift by $\Delta z$ = 0.1. The vertical black dotted line marks the length of the long trough presented in \citet{becker2015}. Middle panel: The effective optical depth integrated over the trough length. The colour of the pixels reflects the number of points in each bin in log scale. Also shown is the 2$\sigma$ limit on the effective optical depth ($\tau_{\rm eff} > 7.2$), measured along the trough of ULAS J0148+0600 from \citet{becker2015} (black arrow). Right panel: A map showing the spatial distribution of trough lengths in our simulation.}
\label{troughs}
\end{figure*}

We show the cumulative distribution function (CDF) of effective optical depths in Figure \ref{opacities} compared with the distribution from \citet{bosman2018}. Here we use the conventional definition $\tau_{\rm eff} = - \log(\bar{F})$, where $\tau_{\rm eff}$ is the effective optical depth and $\bar{F}$ is the mean flux measured in 50 cMpc/$h$ segments of the Ly$\alpha$ forest. We emphasise that the simulations were calibrated to match the observed mean flux, and once this was reproduced the agreement between the observed and simulated $\tau_{\rm eff}$ CDFs came out naturally. Another compilation of effective optical depths is presented in \citet{eilers2018}, however due to systematic differences between the two datasets it is not possible to find a single reionization history that fits both measurements at once. We therefore limit our study here to a comparison with the \citet{bosman2018} compilation. To measure the $\tau_{\rm eff}$ distributions, we create 500 realisations of the \citet{bosman2018} sample and draw different sightlines from our sample in the same redshift distribution. We show the 15$^{\rm th}$ and 85$^{\rm th}$ percentiles of the different CDFs here. As in \citet{kulkarni2019}, in this late reionization model we find good agreement with the observed sample, although we still fail to reproduce the high opacity end of the distribution in the $z=5.2$ redshift bin. This will be discussed later with reference to the incidence rate of long troughs in our simulations.

We present some further results from our simulation in Figure \ref{history}. The top left panel shows the assumed volume emissivity in our simulation compared with what was used in \citet{kulkarni2019}. There are several reasons why we require a slightly different evolution to the assumed emissivity in that work. First, we assume a lower photon energy than in that work to try and recover temperatures that are closer to the Ly$\alpha$ forest measurements. Using this lower photon energy requires us to have a slightly faster reionization to recover the observed mean flux -- since the gas does not reach as high a temperature as it is ionized, we require it to be ionized somewhat later so it has less time to cool down. Second, since we construct the spectra along a lightcone through the simulation, there are  slight changes  to the mean flux we recover from the simulations and therefore we require a different emissivity evolution. Third, as mentioned above, we also would like this simulation to match the mean flux of the Ly$\alpha$ forest below redshift 5. We find that this requires a rapid upturn of the ionizing emissivity just above redshift 5 (which could perhaps be attributed to the onset of an increasing contribution of AGN to the photoionizing background and the onset of \ion{He}{ii} reionization, e.g. \citealt{haardtmadau2012,puchwein2019,kulkarni2019agn}). We further only measure the mean flux in spectra made along a lightcone taken in one direction through our simulation volume. If this exercise was repeated for a larger sample of sightlines, the required emissivity evolution would likely be somewhat different. Furthermore, as already mentioned, the different measurements of the mean flux in this redshift range by \citet{bosman2018} and \citet{eilers2018} are also not fully consistent within the quoted respective errors. We therefore caution against over-interpreting the emissivity evolution presented here. 

Although a decline in the galaxy contribution to the UVB is predicted in models such as \citet{haardtmadau2012} and \citet{puchwein2019}, it is difficult to find a physical interpretation for why the galaxy-like component should drop as rapidly as required here. It is perhaps also concerning that this rapid drop is required just as reionization ends in our simulations. However, the only other way to decrease the mean flux would be if the IGM was much colder, and it would be equally difficult to explain why it should cool so quickly since it needs to be reasonably warm to achieve the observed mean flux at redshift 6. The AGN-like component that we invoke to match the mean flux at lower redshifts becomes dominant over the galaxies much earlier than in other, more physically motivated models \citep[e.g.,][]{puchwein2019} and is in tension with the shape of the UV background inferred from metal lines \citep{boksenberg2003} and measurements of the ionizing emissivity of galaxies \citep{steidel2018} at somewhat lower redshifts. We leave a more careful investigation of the relation of the mean flux of the Ly$\alpha$ forest and the ionizing emissivity to future work, preferably using higher resolution simulations with more careful modelling of the gas temperature, and an improved measurement of the mean flux and the flux PDF.

The other panels of Figure \ref{history} compare our simulation to other properties of the IGM. In this model, the volume is 50 per cent ionized at $z = 6.7$ and 99.9 per cent ionized at $z = 5.2$. In general, the agreement between the model and the observed IGM properties is good. However, as discussed above, since we not self-consistently modelling the QSO-like component of our emissivity, our temperatures below redshift 5 in the voids maybe be underestimated and our photoionization rate will be overestimated. We emphasise though that information from this redshift range is only included as the foregrounds to our  Ly$\beta$ forest modelling and does not affect our main conclusions. The main differences between the model presented here and \citet{kulkarni2019} are driven by the temperature of the gas. Our choice of a lower photon energy here brings the temperature at mean density in the simulations closer to the observations, however as discussed above this is at the expense of introducing a rather sharp drop in the evolution of the emissivity at $z \sim 6$. To achieve the same mean flux, the lower temperatures require a higher photoionization rate. This also seems to be closer to the observed values. We also show for comparison the photoionization rate from the \citet{haardtmadau2012} model, as this was used when running the underlying hydrodynamic simulation. We caution though that the published temperature measurements rely heavily on calibration with numerical simulations that assume a homogeneous UV background, and do not account for the effects of inhomogeneous reionization and spatial fluctuations in the temperature-density relation \citep{furlanetto2009,lidz2014,daloisio2015,keating2018} which is likely to bias the published measurements low.

\section{Long troughs}

\begin{figure*}
  \includegraphics[width=2\columnwidth]{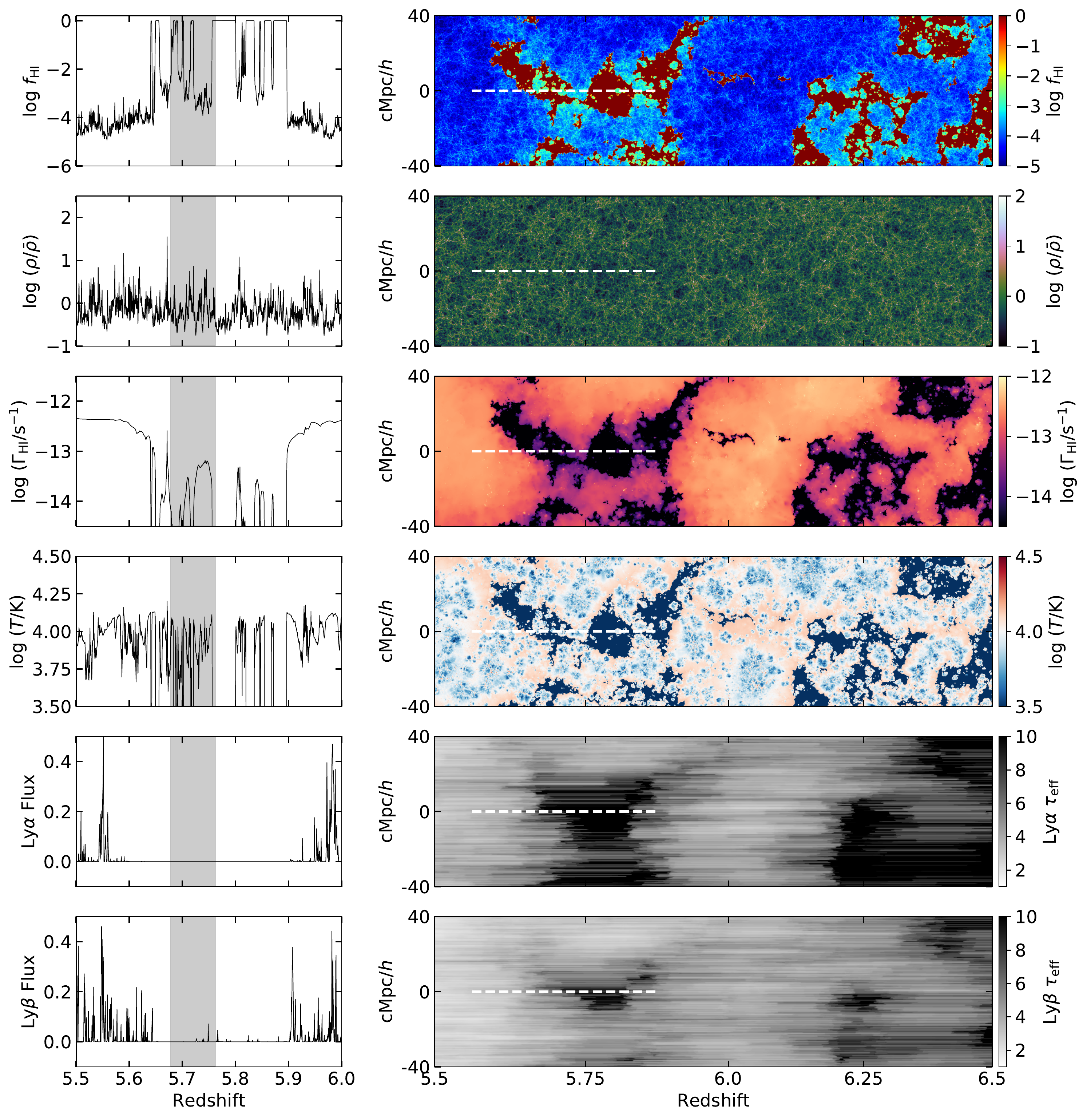}
  \caption{Properties of a sightline containing a long (98.2 cMpc/$h$) trough. The left panel shows a 1D sightline and the right panel shows 2D maps. The white dashed line on the right column corresponds to the redshift range where we measure the Ly$\alpha$ absorption trough. From top to bottom: the neutral hydrogen fraction, gas density, \ion{H}{i} photoionization rate, gas temperature, Ly$\alpha$ flux (left) and  Ly$\alpha$ effective optical depth (right) and Ly$\beta$ flux (left) and  Ly$\beta$ effective optical depth (right) measured over 50 cMpc/$h$ segments. The grey shaded region on the left panels shows the redshift range corresponding to the FWHM of the narrowband filter used to search for LAEs.}
  \label{sightline}
\end{figure*}

We next measure the longest trough along 250,000 sightlines along our extracted lightcone. We post-process the spectra to match the observations, by rebinning onto coarser pixels, convolving with an appropriate instrument profile and adding noise. We measure the length of the troughs by measuring out to a maximum redshift $z_{\rm max} = 5.879$ (the high redshift point of the \citealt{becker2015} trough). We then search for spikes in transmission below this redshift which we assume are regions of the spectrum that show transmission greater than the 1$\sigma$ noise level for at least four consecutive pixels and that have a combined significance of more than 5$\sigma$. We then define the regions in between these transmission spikes as the troughs, and record the longest trough measured for each sightline. As shown in the left panel of Figure \ref{troughs}, our simulation does produce troughs in the Ly$\alpha$ forest as long as the 110 cMpc/$h$ trough of  ULAS J0148+0600, however these are very rare in our simulation (approximately 1 in 10,000). In a recent paper, \citet{giri2019} also presented an investigation of neutral islands at the end of reionization, and noted that they find islands long enough to reproduce the long trough of ULAS J0148+0600. Their simulation volume is much larger (714 cMpc \textit{vs.} 160 cMpc/$h$ in this work) and is therefore more suited for studying the statistics of these long troughs at the end of reionization. Detailed comparisons with observations will however also require that the  simulations are calibrated to match the properties of the Ly$\alpha$ forest, such as in this work.

The long troughs we do find are as dark as measured by \citet{becker2015}, who estimated that the effective optical depth was greater than 7.2 along the absorption trough of  ULAS J0148+0600 (middle panel of Figure \ref{troughs}). We find that troughs with length up to $\sim$ 80 cMpc/$h$ have a relatively high incidence rate in the simulation, but the incidence rate falls rapidly above this point. The most natural explanation for this is that reionization is still ending too early in our simulations. As discussed above, there are 50 cMpc/$h$ segments of the Ly$\alpha$ forest that are observed to be totally opaque at $z=5.2$ which our simulation does not reproduce. If reionization ended later, then the neutral islands would be longer and so would the associated troughs. We demonstrate this by moving the redshift at which we begin measuring the troughs at to a higher redshift ($\Delta z = 0.1$). As shown in the left panel of Figure \ref{troughs}, this has the effect of increasing the length of our simulated troughs and we now find many more longer than 110 cMpc/$h$. However, we note that we were not successful in finding a model where reionization ended later while still matching the observed mean flux. 

\begin{figure*}
  \includegraphics[width=2\columnwidth]{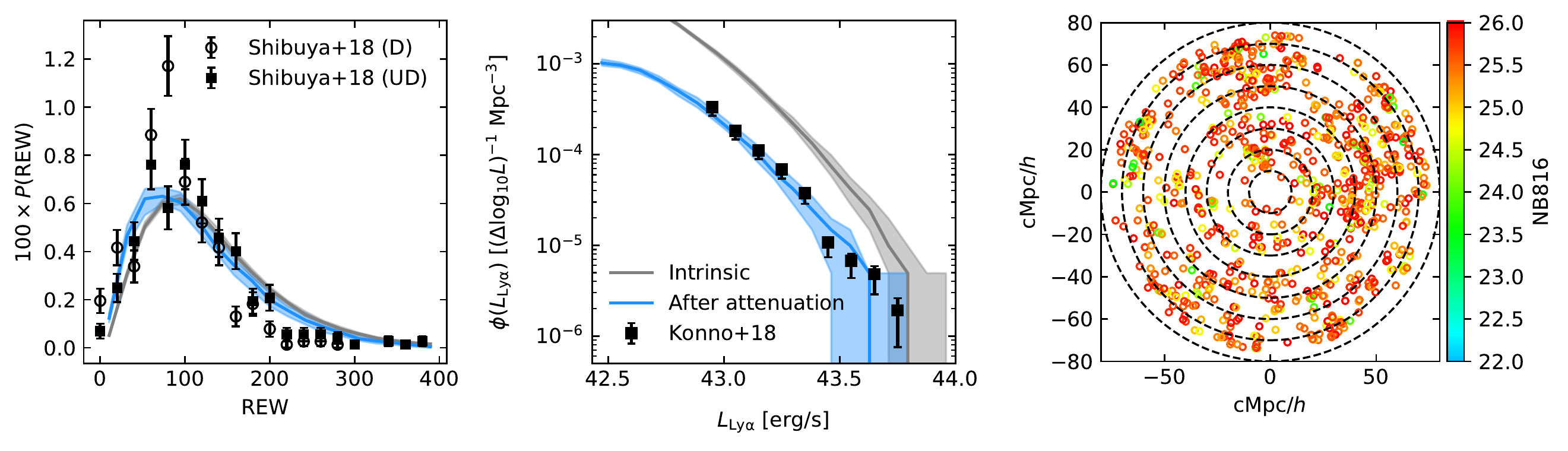}
  \caption{Left: Rest-frame equivalent width distribution of our simulated LAEs (blue line) compared to observations from \citet{shibuya2018} in Deep (D) and UltraDeep (UD) fields. Middle: The LAE luminosity function of our simulated LAEs (blue line) compared to observations from \citet{konno2018}. In both panels, the grey line shows the intrinsic distribution and the blue line shows the distribution after the IGM attenuation is taken into account. The  solid lines are the median values of 50 realisations of the LAE population and the shaded regions represent the 15$^{\rm th}$/85$^{\rm th}$  per cent confidence intervals. Right: The spatial distribution of one realisation of our LAEs centred around the trough shown in Figure \ref{sightline}. The colour of the points represents their NB816 magnitude. The dashed black lines show circles centred around the trough with radii increasing by 10 cMpc/$h$ for each circle.}
  \label{LAEsummary}
\end{figure*}

Another point worth making is that the size of our simulation volume is still small in the context of reionization studies \citep[e.g.,][]{iliev2014}. We therefore do not have a large sample of neutral islands at low redshift in our simulation. This is shown in the right panel of Figure \ref{troughs} -- all of the long troughs are clustered in one section of the box. This means that the deficit of long troughs in our simulations could be caused by an unfavourable geometry in these remaining neutral islands, i.e. they may just happen to not be very extended in the direction in which we extract our lightcone. A stronger statement about the incidence rate of long troughs for a given ionization history will therefore require larger volumes while still maintaining resolution comparable to the simulation presented here. We therefore leave a more detailed study of the incidence rate of these long troughs for future work.

An example of a long (98.2 cMpc/$h$) trough is shown in Figure \ref{sightline}, as well as the associated physical quantities of the gas along the trough. We find that the region spanned by the trough is coincident with an island of neutral hydrogen. It is clear from this figure that the shape of the neutral islands is directly linked to the trough length, and one could imagine that a sightline taken at a small angle to the line of sight shown here would produce a longer trough. As seen in Figure \ref{sightline}, the UVB is also lower in the vicinity of the neutral island, as it is only being illuminated by sources in one direction. This means that the completely neutral gas is surrounded by gas with an \ion{H}{i} fraction of about 10$^{-3}$, still neutral enough to saturate the Ly$\alpha$ transmission. We also find Ly$\beta$ transmission along the trough, as is observed. This can be explained by small pockets of lightly ionized gas within this neutral island. The photoionization rate inside these pockets is still low, however the gas is ionized enough to allow transmission of Ly$\beta$, since Ly$\beta$ has an oscillator strength $\sim 5$ times lower that Ly$\alpha$.

\section{Lyman-$\alpha$ emitters}

A convincing model of the long trough in  ULAS J0148+0600 must also explain the observed deficit of LAEs around that sightline. Since we construct our spectra along a planar lightcone taken at an angle wrapped through our simulation, we cannot make use of periodic boundary conditions to centre a simulated trough in the middle of the lightcone. We therefore pick the sightline we are interested in (in this case, the sightline shown in Figure \ref{sightline}) and rerun the radiative transfer simulation to extract a new lightcone centred on this sightline. This means that our analysis of the LAEs around the trough is restricted to a single sightline. However, since the size of our volume is close to the field of view of HSC and also our longest troughs are co-spatial (Figure \ref{troughs}) an analysis of more than one trough in the simulation would not really remove the effects of cosmic variance. We chose this trough because, although it is slightly shorter than the trough of ULAS J0148+0600, it showed Ly$\beta$ emission with similar properties to the observed long trough, which shows only one Ly$\beta$ transmission spike in the redshift range covered by the narrowband filter used in \citet{becker2018}.

\begin{figure*}
  \includegraphics[width=2\columnwidth]{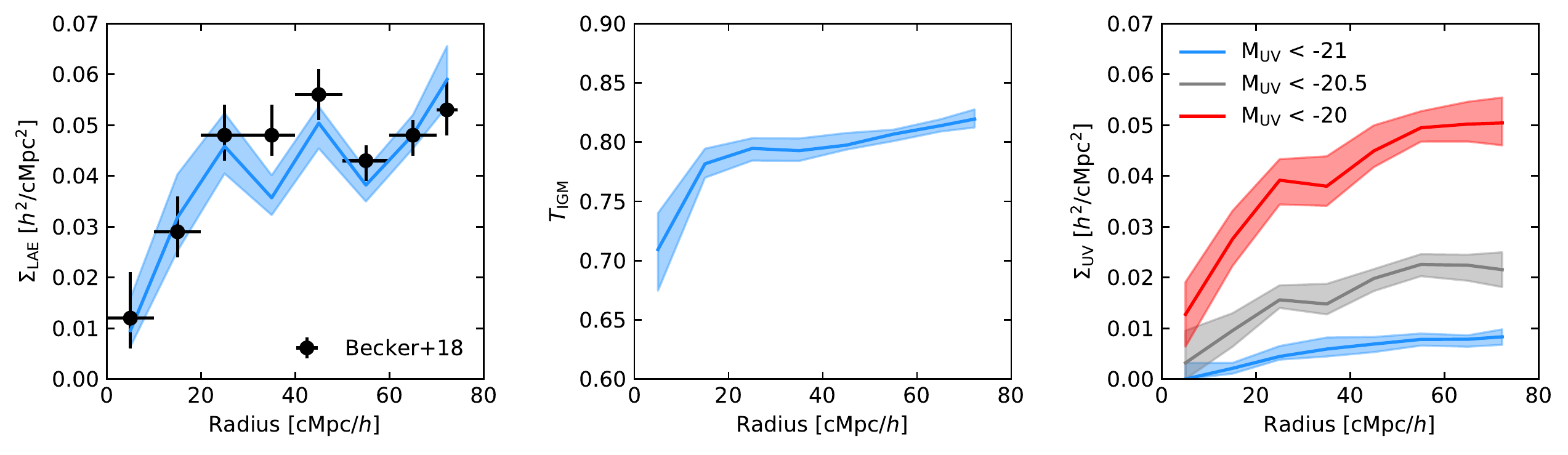}
  \caption{Left: The surface density of LAEs measured in annuli centred around the trough. The black points are the \citet{becker2018} blue line is the median surface density measured in our LAE realisation. The shaded regions represent the 15$^{\rm th}$/85$^{\rm th}$   per cent confidence intervals. Middle: The averaged IGM transmission measured in annuli centred around the trough, which we define as the ratio of the observed and intrinsic LAE luminosities. Right: The surface density of LBGs with $M_{\rm UV} < -20$ (red), $M_{\rm UV} < -20.5$ (grey) and $M_{\rm UV} < -21$ (blue) around the same sightline.}
  \label{LAEsd}
\end{figure*}

To model the LAEs around our simulated trough, we follow the method described in detail in \citet{weinberger2019}. We note that this method is similar to the LAE modelling in \citet{davies2018} and \cite{becker2018}, although as we will describe later we also take the effect of the local IGM opacity into account. We first assign UV luminosities to the dark matter haloes by abundance matching to the UV luminosity function at redshift 5.9 \citep{bouwens2015}. We assume a 50 Myr duty cycle such that only a fraction of our galaxies are UV bright at a given time, reflecting the bursty nature of star formation \citep{trenti2010}. Following \citet{dijkstra2012}, we then assign a rest-frame equivalent width of Ly$\alpha$ emission to each halo based on its UV luminosity using the relation
\begin{equation}
P({\rm REW} | M_{\rm UV}) \propto \exp \left( \frac{- {\rm REW}}{{\rm REW}_{\rm c}(M_{\rm UV})} \right),
\end{equation}
where REW is the rest-frame equivalent width, $M_{\rm UV}$ is the UV magnitude and REW$_{\rm c}$ is defined as
\begin{equation}
{\rm REW}_{\rm c}(M_{\rm UV}) = 23 + 7(M_{\rm UV} + 21.9) + 6(z-4).
\end{equation}
The normalization is defined such that the population has ${\rm REW}_{\rm min} \leq {\rm REW} \leq {\rm REW}_{\rm max}$, where we choose ${\rm REW}_{\rm max} = 300 \, {\rm \AA}$. The minimum REW is defined in terms of $M_{\rm UV}$ as
\begin{equation}
{\rm REW}_{\rm min} =
\begin{cases}
-20 \, {\rm \AA} & M_{\rm UV} < -21.5,\\
17.5 \, {\rm \AA} & M_{\rm UV} > -19.0,\\
-20 + 6(M_{\rm UV} + 21.5)^2 \, {\rm \AA} & {\rm otherwise}.
\end{cases}
\end{equation}
We note that the assignment of  UV magnitude and REW is a random process, since only a fraction of the haloes are UV bright at a given time and the REW of each halo is drawn from a distribution. We therefore repeated this process 50 times, to generate 50 different realisations of LAE populations.

We next accounted for the attenuation of Ly$\alpha$ emission due to neutral gas around the haloes. The default resolution of our radiative transfer simulations is too coarse to resolve the circumgalactic medium (CGM) of halos, so we re-extract a higher resolution skewer of gas density and peculiar velocity in a 10 cMpc/$h$ radius around each halo. The resolution of these skewers is 9.7 ckpc/$h$. We insert this short high resolution sightline into the longer lower resolution sightline to model the attenuation of Ly$\alpha$ emission due to both the CGM and IGM. Since we do not know the ionization state of the gas in the high resolution sightline, we assume that the gas is in ionization equilibrium calculated using the photoionization rates from our radiative transfer simulation plus the self-shielding prescription of \citet{chardin2018}. We then calculate the optical depth along each sightline.

We construct mock LAE spectra by using the UV magnitude to get the flux at rest-frame 1600 \AA, and assuming that $F_{\lambda} \propto \lambda^{-2}$. To model the Ly$\alpha$ emission, we add a gaussian emission line with a REW given by the prescription outlined above. As in \citet{weinberger2019}, we also assume some velocity shift between the redshift of the host halo and the redshift of the Ly$\alpha$ emission, which accounts for the complex radiative transfer of the Ly$\alpha$ line out of the halo. We take this to be a constant factor times the circular velocity of the host halo, with the constant factor chosen to ensure good agreement with the rest-frame equivalent widths and luminosities of observed LAEs. We then attenuate the spectrum by multiplying by the normalised Ly$\alpha$ transmission we calculate along each sightline. In this work we found that a velocity shift of $1.4 v_{\rm circ}$ was a good choice. Using these spectra, we then measure the magnitude of the LAEs in the $i2$ and $NB816$ bands and select LAEs that have $NB816 \leq 26.0$ and $i2 - NB816 \geq 1.2$ as in \citet{becker2018}. We note that \citet{becker2018} also use $r2$ band magnitudes as part of their selection process, but we can not model that band here because our spectra do not extend down to sufficiently low redshift. We calculate the luminosity and rest-frame equivalent width for each LAE by integrating over our constructed LAE spectra. As in \citet{weinberger2019}, the agreement between the observed and simulated LAEs is very good, i.e. our modelled LAE population agrees with the observed LAE luminosity function and rest-frame equivalent width distribution from the SILVERRUSH survey (Figure \ref{LAEsummary}), however we note that our box is too small to reproduce the brightest LAEs (seen by the rapid cutoff of our modelled luminosity function).

\begin{figure*}
  \includegraphics[width=2\columnwidth]{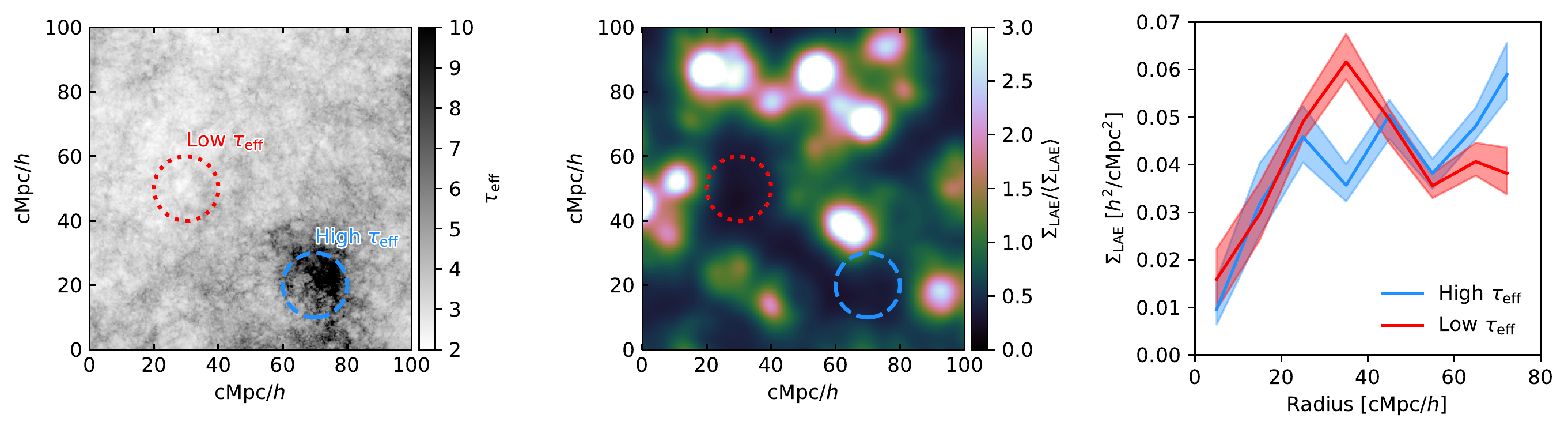}
  \caption{Left: Map of the effective optical depth measured in a 50 cMpc/$h$ segment centred on the peak transmission wavelength of the narrowband filter. The red and blue circles mark out regions that have low and high effective optical depths. Middle: Smoothed average LAE surface density map for the same field as in the right panel.  Right: Surface density of LAEs for the low and high effective optical depth regions.}
  \label{taueff_LAEs}
\end{figure*}

We next compare the spatial distribution of LAEs around our simulated trough (right panel of Figure \ref{LAEsummary}). We show here the projected position of LAEs coloured by their  $NB816$ magnitude. As in \citet{becker2018}, we find a large deficit of LAEs in the immediate vicinity of the trough. This is to be expected, since the low density regions will be the last to ionize. We therefore expect our remaining neutral islands to trace underdense regions of the IGM. Furthermore, since the Ly$\alpha$ optical depth along the trough is high, we also expect the attenuation of Ly$\alpha$ emission to be stronger in the region around the trough. In the left panel of Figure \ref{LAEsd}, we show the LAE surface density computed in annuli centred around the trough. The results from our simulated LAE are shown in blue (again, the shaded regions mark the 15$^{\rm th}$/85$^{\rm th}$ percentiles of our 50 different LAE realisations and the solid line is the median value). The results from \citet{becker2018} are shown in black. Similar to the observations, we find that the surface density of LAEs begins to decline rapidly in the inner 30 cMpc/$h$ around the trough. This is predominantly due to the trough passing through an underdense region, although we note that that this effect is complemented by a decrease in the IGM transmission fraction of about 10 per cent close to the trough (middle panel of Figure \ref{LAEsd}). Here we define the IGM transmission fraction as $T_{\rm IGM} = L_{\rm obs}$/$L_{\rm int}$ where $L$ is the observed/intrinsic LAE luminosity \citep{mesinger2015}. This lower transmission fraction is in agreement with \citet{sadoun2017}, who suggest that the observed number of LAEs will be lower in regions with a lower photoionizing background, due to the increased filling factor of self-shielded systems in the CGM of the host galaxies. However, it does not appear to be the dominant effect here.

In the right panel of Figure \ref{LAEsd}, we show the surface density of UV-selected galaxies that have $M_{\rm UV} < -20$, $-20.5$ and $-21$. For the cut with $M_{\rm UV} < -20$, we find that the surface density far from the trough is well matched to the LAEs. However, the decline in surface density is even stronger for this selection, already beginning to decline at a radius 50 cMpc/$h$ from the trough and highlighting that this sightline passes through an underdense region. The median mass of the LAEs that pass the colour cuts used here is $4.1 \times 10^{10} \, M_{\odot}$/$h$. For the UV selected galaxies, the median masses are $(1.1,1.7,3.1) \times 10^{11} \, M_{\odot}$/$h$ for $M_{\rm UV} < (-20, -20.5, -21)$.

\section{Comparison with other work}
\label{sec:comparision}

As shown in \citet{becker2018}, the fluctuating UVB model presented in \citet{davies2016} can also reproduce the deficit of LAEs in high $\tau_{\rm eff}$ regions. In this case, a short mean free path leads to large fluctuations in the UVB. This produces regions that, while still ionized, have a \ion{H}{i} fraction high enough to saturate transmission of Ly$\alpha$. We do not find that this is the case in our simulations, as once the ionized regions have percolated the mean free path becomes long and any fluctuations in the UVB are quickly washed away. Our late reionization model is still similar to the  \citet{davies2016} in that it is the underdense regions that produce the high $\tau_{\rm eff}$ sightlines. In our case, however, this is because reionization has not yet ended and the voids host the residual neutral islands. In both models, the underdense regions host the high $\tau_{\rm eff}$ sightlines because they are far from the regions where the space density of ionizing sources is highest. Linking the \citet{davies2016} model to a late reionization scenario removes the main issues with that explanation. First, that the mean free path needs to become much shorter at $z > 5$ than observed by \citet{worseck2014} at redshift 5. In our model, the mean free path does decrease sharply above redshift 5 due to the residual islands of neutral hydrogen. Second, there is no obvious way to predict the redshift evolution of the mean free path in the \citet{davies2016} model -- at each redshift it must be tuned to match constraints from the Ly$\alpha$ forest. We also tune the emissivity in our model to match the observed mean flux, however once this is fixed then the reionization history of our preferred model can be used to make predictions for higher redshifts. In this sense, our late reionization model can be thought of as a dynamic version of the \citet{davies2016} model.

Ideally one would still like to discriminate between the two models though, to answer if indeed the high $\tau_{\rm eff}$ regions are mildly ionized or completely neutral. One difference between the two scenarios is that the fluctuating UVB model predicts that the underdense regions will always coincide with high $\tau_{\rm eff}$ sightlines, and the low $\tau_{\rm eff}$ sightlines will lie in overdense regions \citep{davies2018}. In a late reionization model though, the underdense regions will align with  high $\tau_{\rm eff}$ sightlines only in the time before they have been ionized. Once the neutral islands have vanished, the voids should instead correspond to the lowest $\tau_{\rm eff}$ regions, as most of the  Ly$\alpha$ forest transmission is dominated by the voids. These recently ionized regions should also be very hot, suppressing recombinations and further enhancing transmission \citep{daloisio2015}. 

We present an example of two such sightlines in Figure \ref{taueff_LAEs}. The left panel shows a map of $\tau_{\rm eff}$ measured in 50 cMpc/$h$ segments centred on the redshift corresponding to peak transmission of the $NB816$ filter. The blue and red circles mark two regions of high and low $\tau_{\rm eff}$. In the middle panel, we show the mean surface density of LAEs in that region. The surface density is taken to be inversely proportional to the square of the distance to the tenth nearest neighbour and the map has been smoothed with a gaussian kernel. The red and blue circles correspond to the same regions as in the left panel. It is clear that both of these regions show a deficit of LAEs, suggesting that both of these are underdense regions. This is shown explicitly in the right panel, where we compare the surface density of LAEs as a function of radius for the two lines of sight. In the \citet{davies2016} model, you would instead expect that the low $\tau_{\rm eff}$ region should show an overdensity of LAEs. This suggests that future observations searching for LAEs around sightlines that show significant Ly$\alpha$ forest transmission may be interesting for differentiating between these two models.

\section{Summary and Conclusions}

We have shown here that long Gunn-Peterson troughs observed in the Ly$\alpha$ forest above redshift 5 can be explained by islands of neutral hydrogen in the late reionization scenario presented in \citet{kulkarni2019}. We have demonstrated that this model also explains the significant deficit of LAEs detected around these high opacity sightlines by \citet{becker2018}. While these observations can also be explained by a model in which the amplitude of the UVB is low in underdense regions, we argue that the model presented here is more consistent with the observed evolution of the mean free path for ionizing photons. We further present predictions for the distribution of LAEs around the low opacity sightlines which should allow future observations to discriminate between these two scenarios. This however will still not be a definitive test of neutral gas in the IGM, so other tests should also be pursued \citep{lidz2015}.

We note that the reionization model presented here still does not reproduce the most opaque sightlines in the range $5.1 < z < 5.3$ within the 15$^{\rm th}$/85$^{\rm th}$ interval. Searching for more saturated sightlines at low redshifts would be useful in constraining whether these sightlines are rare. If they turn out not to be outliers, this may call for a reionization history that ends even later than shown here. This would also likely increase the incidence rate of long ($>$ 100 cMpc/$h$) troughs. We note also that the temperature of the IGM is still a significant uncertainty in our models, as it can have a large effect on the mean flux (as the recombination rate and thus the effective optical depth for a given UVB amplitude depend rather strongly on temperature). More accurate models of the ionization history of the IGM would benefit from more estimates of the temperature at high redshift, perhaps also using methods that account for the pressure effects for spatial fluctuations of the temperature-density relation due to reionization \citep[e.g.,][Puchwein et al. in prep]{onorbe2019}.

As more high signal-to-noise Ly$\alpha$ forest sightlines are accumulated and measurements of the mean flux and flux PDF further improve, there will be benefit in moving from studies of integrated quantities such as the effective optical depth to more detailed studies of individual features. Future work modelling the occurrence rate and distribution of these troughs, as well as the transmission peaks between them, should provide more detailed constraints on the latter half of reionization \citep{chardin2017spikes,garaldi2019spikes}. Tying these Ly$\alpha$ forest measurements to the galaxy population may also yield additional information such as the sources that contribute to reionization \citep{kakiichi2018}, especially once \textit{JWST} is operational. Studies utilising also the Ly$\beta$ forest should allow to push constraints on the reionization history  outside of the immediate vicinity of high redshift quasars to $z \sim 6.5$ and beyond.

\section*{Acknowledgements}

We thank George Becker for useful discussions and for the suggestion that motivated Section \ref{sec:comparision}. This work was performed using resources provided by the Cambridge Service for Data Driven Discovery (CSD3) operated by the University of Cambridge Research Computing Service (www.csd3.cam.ac.uk), provided by Dell EMC and Intel using Tier-2 funding from the Engineering and Physical Sciences Research Council (capital grant EP/P020259/1), and DiRAC funding from the Science and Technology Facilities Council (www.dirac.ac.uk). This work used the COSMA Data Centric system at Durham University, operated by the Institute for Computational Cosmology on behalf of the STFC DiRAC HPC Facility. This equipment was funded by a BIS National E-infrastructure capital grant ST/K00042X/1, DiRAC Operations grant ST/K003267/1 and Durham University. DiRAC is part of the National E- Infrastructure. This work was supported by the ERC Advanced Grant 320596 ``The Emergence of Structure During the Epoch of Reionization''.

\bibliographystyle{mnras} \bibliography{/data/vault/lck35/ref}

\begin{thebibliography}{}
\makeatletter
\relax
\def\mn@urlcharsother{\let\do\@makeother \do\$\do\&\do\#\do\^\do\_\do\%\do\~}
\def\mn@doi{\begingroup\mn@urlcharsother \@ifnextchar [ {\mn@doi@}
  {\mn@doi@[]}}
\def\mn@doi@[#1]#2{\def\@tempa{#1}\ifx\@tempa\@empty \href
  {http://dx.doi.org/#2} {doi:#2}\else \href {http://dx.doi.org/#2} {#1}\fi
  \endgroup}
\def\mn@eprint#1#2{\mn@eprint@#1:#2::\@nil}
\def\mn@eprint@arXiv#1{\href {http://arxiv.org/abs/#1} {{\tt arXiv:#1}}}
\def\mn@eprint@dblp#1{\href {http://dblp.uni-trier.de/rec/bibtex/#1.xml}
  {dblp:#1}}
\def\mn@eprint@#1:#2:#3:#4\@nil{\def\@tempa {#1}\def\@tempb {#2}\def\@tempc
  {#3}\ifx \@tempc \@empty \let \@tempc \@tempb \let \@tempb \@tempa \fi \ifx
  \@tempb \@empty \def\@tempb {arXiv}\fi \@ifundefined
  {mn@eprint@\@tempb}{\@tempb:\@tempc}{\expandafter \expandafter \csname
  mn@eprint@\@tempb\endcsname \expandafter{\@tempc}}}

\bibitem[\protect\citeauthoryear{{Aubert} \& {Teyssier}}{{Aubert} \&
  {Teyssier}}{2008}]{aubert2008}
{Aubert} D.,  {Teyssier} R.,  2008, \mn@doi [\mnras]
  {10.1111/j.1365-2966.2008.13223.x}, \href
  {http://adsabs.harvard.edu/abs/2008MNRAS.387..295A} {387, 295}

\bibitem[\protect\citeauthoryear{{Aubert} \& {Teyssier}}{{Aubert} \&
  {Teyssier}}{2010}]{aubert2010}
{Aubert} D.,  {Teyssier} R.,  2010, \mn@doi [\apj]
  {10.1088/0004-637X/724/1/244}, \href
  {https://ui.adsabs.harvard.edu/abs/2010ApJ...724..244A} {724, 244}

\bibitem[\protect\citeauthoryear{{Becker} \& {Bolton}}{{Becker} \&
  {Bolton}}{2013}]{becker2013}
{Becker} G.~D.,  {Bolton} J.~S.,  2013, \mn@doi [\mnras]
  {10.1093/mnras/stt1610}, \href
  {http://adsabs.harvard.edu/abs/2013MNRAS.436.1023B} {436, 1023}

\bibitem[\protect\citeauthoryear{{Becker}, {Bolton}, {Haehnelt}  \&
  {Sargent}}{{Becker} et~al.}{2011}]{becker2011temp}
{Becker} G.~D.,  {Bolton} J.~S.,  {Haehnelt} M.~G.,   {Sargent} W.~L.~W.,
  2011, \mn@doi [\mnras] {10.1111/j.1365-2966.2010.17507.x}, \href
  {http://adsabs.harvard.edu/abs/2011MNRAS.410.1096B} {410, 1096}

\bibitem[\protect\citeauthoryear{{Becker}, {Bolton}, {Madau}, {Pettini},
  {Ryan-Weber}  \& {Venemans}}{{Becker} et~al.}{2015}]{becker2015}
{Becker} G.~D.,  {Bolton} J.~S.,  {Madau} P.,  {Pettini} M.,  {Ryan-Weber}
  E.~V.,   {Venemans} B.~P.,  2015, \mn@doi [\mnras] {10.1093/mnras/stu2646},
  \href {http://adsabs.harvard.edu/abs/2015MNRAS.447.3402B} {447, 3402}

\bibitem[\protect\citeauthoryear{{Becker}, {Davies}, {Furlanetto}, {Malkan},
  {Boera}  \& {Douglass}}{{Becker} et~al.}{2018}]{becker2018}
{Becker} G.~D.,  {Davies} F.~B.,  {Furlanetto} S.~R.,  {Malkan} M.~A.,  {Boera}
  E.,   {Douglass} C.,  2018, \mn@doi [\apj] {10.3847/1538-4357/aacc73}, \href
  {http://adsabs.harvard.edu/abs/2018ApJ...863...92B} {863, 92}

\bibitem[\protect\citeauthoryear{{Boera}, {Becker}, {Bolton}  \&
  {Nasir}}{{Boera} et~al.}{2019}]{boera2019}
{Boera} E.,  {Becker} G.~D.,  {Bolton} J.~S.,   {Nasir} F.,  2019, \mn@doi
  [\apj] {10.3847/1538-4357/aafee4}, \href
  {http://adsabs.harvard.edu/abs/2019ApJ...872..101B} {872, 101}

\bibitem[\protect\citeauthoryear{{Boksenberg}, {Sargent}  \&
  {Rauch}}{{Boksenberg} et~al.}{2003}]{boksenberg2003}
{Boksenberg} A.,  {Sargent} W.~L.~W.,   {Rauch} M.,  2003, arXiv:0307557, \href
  {http://adsabs.harvard.edu/abs/2003astro.ph..7557B} {}

\bibitem[\protect\citeauthoryear{{Bolton}, {Becker}, {Raskutti}, {Wyithe},
  {Haehnelt}  \& {Sargent}}{{Bolton} et~al.}{2012}]{bolton2012}
{Bolton} J.~S.,  {Becker} G.~D.,  {Raskutti} S.,  {Wyithe} J.~S.~B.,
  {Haehnelt} M.~G.,   {Sargent} W.~L.~W.,  2012, \mn@doi [\mnras]
  {10.1111/j.1365-2966.2011.19929.x}, \href
  {http://adsabs.harvard.edu/abs/2012MNRAS.419.2880B} {419, 2880}

\bibitem[\protect\citeauthoryear{{Bolton}, {Puchwein}, {Sijacki}, {Haehnelt},
  {Kim}, {Meiksin}, {Regan}  \& {Viel}}{{Bolton} et~al.}{2017}]{bolton2017}
{Bolton} J.~S.,  {Puchwein} E.,  {Sijacki} D.,  {Haehnelt} M.~G.,  {Kim} T.-S.,
   {Meiksin} A.,  {Regan} J.~A.,   {Viel} M.,  2017, \mn@doi [\mnras]
  {10.1093/mnras/stw2397}, \href
  {http://adsabs.harvard.edu/abs/2017MNRAS.464..897B} {464, 897}

\bibitem[\protect\citeauthoryear{{Bosman}, {Fan}, {Jiang}, {Reed}, {Matsuoka},
  {Becker}  \& {Haehnelt}}{{Bosman} et~al.}{2018}]{bosman2018}
{Bosman} S.~E.~I.,  {Fan} X.,  {Jiang} L.,  {Reed} S.,  {Matsuoka} Y.,
  {Becker} G.,   {Haehnelt} M.,  2018, \mn@doi [\mnras]
  {10.1093/mnras/sty1344}, \href
  {http://adsabs.harvard.edu/abs/2018MNRAS.479.1055B} {479, 1055}

\bibitem[\protect\citeauthoryear{{Bouwens}, {Illingworth}, {Oesch}, {Caruana},
  {Holwerda}, {Smit}  \& {Wilkins}}{{Bouwens} et~al.}{2015}]{bouwens2015}
{Bouwens} R.~J.,  {Illingworth} G.~D.,  {Oesch} P.~A.,  {Caruana} J.,
  {Holwerda} B.,  {Smit} R.,   {Wilkins} S.,  2015, \mn@doi [\apj]
  {10.1088/0004-637X/811/2/140}, \href
  {http://adsabs.harvard.edu/abs/2015ApJ...811..140B} {811, 140}

\bibitem[\protect\citeauthoryear{{Calverley}, {Becker}, {Haehnelt}  \&
  {Bolton}}{{Calverley} et~al.}{2011}]{calverley2011}
{Calverley} A.~P.,  {Becker} G.~D.,  {Haehnelt} M.~G.,   {Bolton} J.~S.,  2011,
  \mn@doi [\mnras] {10.1111/j.1365-2966.2010.18072.x}, \href
  {http://adsabs.harvard.edu/abs/2011MNRAS.412.2543C} {412, 2543}

\bibitem[\protect\citeauthoryear{{Cen}}{{Cen}}{1992}]{cen1992}
{Cen} R.,  1992, \mn@doi [\apjs] {10.1086/191630}, \href
  {http://adsabs.harvard.edu/abs/1992ApJS...78..341C} {78, 341}

\bibitem[\protect\citeauthoryear{{Chardin}, {Haehnelt}, {Aubert}  \&
  {Puchwein}}{{Chardin} et~al.}{2015}]{chardin2015}
{Chardin} J.,  {Haehnelt} M.~G.,  {Aubert} D.,   {Puchwein} E.,  2015, \mn@doi
  [\mnras] {10.1093/mnras/stv1786}, \href
  {http://adsabs.harvard.edu/abs/2015MNRAS.453.2943C} {453, 2943}

\bibitem[\protect\citeauthoryear{{Chardin}, {Puchwein}  \&
  {Haehnelt}}{{Chardin} et~al.}{2017}]{chardin2017}
{Chardin} J.,  {Puchwein} E.,   {Haehnelt} M.~G.,  2017, \mn@doi [\mnras]
  {10.1093/mnras/stw2943}, \href
  {http://adsabs.harvard.edu/abs/2017MNRAS.465.3429C} {465, 3429}

\bibitem[\protect\citeauthoryear{{Chardin}, {Haehnelt}, {Bosman}  \&
  {Puchwein}}{{Chardin} et~al.}{2018a}]{chardin2017spikes}
{Chardin} J.,  {Haehnelt} M.~G.,  {Bosman} S.~E.~I.,   {Puchwein} E.,  2018a,
  \mn@doi [\mnras] {10.1093/mnras/stx2362}, \href
  {http://adsabs.harvard.edu/abs/2018MNRAS.473..765C} {473, 765}

\bibitem[\protect\citeauthoryear{{Chardin}, {Kulkarni}  \&
  {Haehnelt}}{{Chardin} et~al.}{2018b}]{chardin2018}
{Chardin} J.,  {Kulkarni} G.,   {Haehnelt} M.~G.,  2018b, \mn@doi [\mnras]
  {10.1093/mnras/sty992}, \href
  {http://adsabs.harvard.edu/abs/2018MNRAS.478.1065C} {478, 1065}

\bibitem[\protect\citeauthoryear{{D'Aloisio}, {McQuinn}  \& {Trac}}{{D'Aloisio}
  et~al.}{2015}]{daloisio2015}
{D'Aloisio} A.,  {McQuinn} M.,   {Trac} H.,  2015, \mn@doi [\apjl]
  {10.1088/2041-8205/813/2/L38}, \href
  {http://adsabs.harvard.edu/abs/2015ApJ...813L..38D} {813, L38}

\bibitem[\protect\citeauthoryear{{D'Aloisio}, {Upton Sanderbeck}, {McQuinn},
  {Trac}  \& {Shapiro}}{{D'Aloisio} et~al.}{2017}]{daloisio2017}
{D'Aloisio} A.,  {Upton Sanderbeck} P.~R.,  {McQuinn} M.,  {Trac} H.,
  {Shapiro} P.~R.,  2017, \mn@doi [\mnras] {10.1093/mnras/stx711}, \href
  {http://adsabs.harvard.edu/abs/2017MNRAS.468.4691D} {468, 4691}

\bibitem[\protect\citeauthoryear{{D'Aloisio}, {McQuinn}, {Davies}  \&
  {Furlanetto}}{{D'Aloisio} et~al.}{2018}]{daloisio2018}
{D'Aloisio} A.,  {McQuinn} M.,  {Davies} F.~B.,   {Furlanetto} S.~R.,  2018,
  \mn@doi [\mnras] {10.1093/mnras/stx2341}, \href
  {http://adsabs.harvard.edu/abs/2018MNRAS.473..560D} {473, 560}

\bibitem[\protect\citeauthoryear{{Davies} \& {Furlanetto}}{{Davies} \&
  {Furlanetto}}{2016}]{davies2016}
{Davies} F.~B.,  {Furlanetto} S.~R.,  2016, \mn@doi [\mnras]
  {10.1093/mnras/stw931}, \href
  {http://adsabs.harvard.edu/abs/2016MNRAS.460.1328D} {460, 1328}

\bibitem[\protect\citeauthoryear{{Davies}, {Becker}  \& {Furlanetto}}{{Davies}
  et~al.}{2018}]{davies2018}
{Davies} F.~B.,  {Becker} G.~D.,   {Furlanetto} S.~R.,  2018, \mn@doi [\apj]
  {10.3847/1538-4357/aac2d6}, \href
  {http://adsabs.harvard.edu/abs/2018ApJ...860..155D} {860, 155}

\bibitem[\protect\citeauthoryear{{Dijkstra} \& {Wyithe}}{{Dijkstra} \&
  {Wyithe}}{2012}]{dijkstra2012}
{Dijkstra} M.,  {Wyithe} J.~S.~B.,  2012, \mn@doi [\mnras]
  {10.1111/j.1365-2966.2011.19958.x}, \href
  {http://adsabs.harvard.edu/abs/2012MNRAS.419.3181D} {419, 3181}

\bibitem[\protect\citeauthoryear{{Eilers}, {Davies}  \& {Hennawi}}{{Eilers}
  et~al.}{2018}]{eilers2018}
{Eilers} A.-C.,  {Davies} F.~B.,   {Hennawi} J.~F.,  2018, \mn@doi [\apj]
  {10.3847/1538-4357/aad4fd}, \href
  {http://adsabs.harvard.edu/abs/2018ApJ...864...53E} {864, 53}

\bibitem[\protect\citeauthoryear{{Fan} et~al.,}{{Fan} et~al.}{2006}]{fan2006}
{Fan} X.,  et~al., 2006, \mn@doi [\aj] {10.1086/504836}, \href
  {http://adsabs.harvard.edu/abs/2006AJ....132..117F} {132, 117}

\bibitem[\protect\citeauthoryear{{Furlanetto} \& {Oh}}{{Furlanetto} \&
  {Oh}}{2009}]{furlanetto2009}
{Furlanetto} S.~R.,  {Oh} S.~P.,  2009, \mn@doi [\apj]
  {10.1088/0004-637X/701/1/94}, \href
  {http://adsabs.harvard.edu/abs/2009ApJ...701...94F} {701, 94}

\bibitem[\protect\citeauthoryear{{Garaldi}, {Compostella}  \&
  {Porciani}}{{Garaldi} et~al.}{2019a}]{garaldi2019}
{Garaldi} E.,  {Compostella} M.,   {Porciani} C.,  2019a, \mn@doi [\mnras]
  {10.1093/mnras/sty3414}, \href
  {http://adsabs.harvard.edu/abs/2019MNRAS.483.5301G} {483, 5301}

\bibitem[\protect\citeauthoryear{{Garaldi}, {Gnedin}  \& {Madau}}{{Garaldi}
  et~al.}{2019b}]{garaldi2019spikes}
{Garaldi} E.,  {Gnedin} N.,   {Madau} P.,  2019b, \mn@doi [\apj]
  {10.3847/1538-4357/ab12dc}, \href
  {https://ui.adsabs.harvard.edu/abs/2019ApJ...876...31G} {876, 31}

\bibitem[\protect\citeauthoryear{{Giri}, {Mellema}, {Aldheimer}, {Dixon}  \&
  {Iliev}}{{Giri} et~al.}{2019}]{giri2019}
{Giri} S.~K.,  {Mellema} G.,  {Aldheimer} T.,  {Dixon} K.~L.,   {Iliev} I.~T.,
  2019, \mn@doi [Monthly Notices of the Royal Astronomical Society]
  {10.1093/mnras/stz2224}, \href
  {https://ui.adsabs.harvard.edu/abs/2019MNRAS.tmp.2192G} {p.~2192}

\bibitem[\protect\citeauthoryear{{Gnedin} \& {Abel}}{{Gnedin} \&
  {Abel}}{2001}]{gnedin2001}
{Gnedin} N.~Y.,  {Abel} T.,  2001, \mn@doi [\na]
  {10.1016/S1384-1076(01)00068-9}, \href
  {https://ui.adsabs.harvard.edu/abs/2001NewA....6..437G} {6, 437}

\bibitem[\protect\citeauthoryear{{Greig}, {Mesinger}, {Haiman}  \&
  {Simcoe}}{{Greig} et~al.}{2017}]{greig2017}
{Greig} B.,  {Mesinger} A.,  {Haiman} Z.,   {Simcoe} R.~A.,  2017, \mn@doi
  [\mnras] {10.1093/mnras/stw3351}, \href
  {http://adsabs.harvard.edu/abs/2017MNRAS.466.4239G} {466, 4239}

\bibitem[\protect\citeauthoryear{{Greig}, {Mesinger}  \& {Ba{\~n}ados}}{{Greig}
  et~al.}{2019}]{greig2019}
{Greig} B.,  {Mesinger} A.,   {Ba{\~n}ados} E.,  2019, \mn@doi [\mnras]
  {10.1093/mnras/stz230}, \href
  {http://adsabs.harvard.edu/abs/2019MNRAS.484.5094G} {484, 5094}

\bibitem[\protect\citeauthoryear{{Haardt} \& {Madau}}{{Haardt} \&
  {Madau}}{2012}]{haardtmadau2012}
{Haardt} F.,  {Madau} P.,  2012, \mn@doi [\apj] {10.1088/0004-637X/746/2/125},
  \href {http://adsabs.harvard.edu/abs/2012ApJ...746..125H} {746, 125}

\bibitem[\protect\citeauthoryear{{Haiman}, {Thoul}  \& {Loeb}}{{Haiman}
  et~al.}{1996}]{haiman1996}
{Haiman} Z.,  {Thoul} A.~A.,   {Loeb} A.,  1996, \mn@doi [\apj]
  {10.1086/177343}, \href
  {https://ui.adsabs.harvard.edu/abs/1996ApJ...464..523H} {464, 523}

\bibitem[\protect\citeauthoryear{{Hui} \& {Gnedin}}{{Hui} \&
  {Gnedin}}{1997}]{huignedin1997}
{Hui} L.,  {Gnedin} N.~Y.,  1997, \mnras, \href
  {http://adsabs.harvard.edu/abs/1997MNRAS.292...27H} {292, 27}

\bibitem[\protect\citeauthoryear{{Iliev}, {Mellema}, {Pen}, {Merz}, {Shapiro}
  \& {Alvarez}}{{Iliev} et~al.}{2006}]{iliev2006b}
{Iliev} I.~T.,  {Mellema} G.,  {Pen} U.-L.,  {Merz} H.,  {Shapiro} P.~R.,
  {Alvarez} M.~A.,  2006, \mn@doi [\mnras] {10.1111/j.1365-2966.2006.10502.x},
  \href {http://adsabs.harvard.edu/abs/2006MNRAS.369.1625I} {369, 1625}

\bibitem[\protect\citeauthoryear{{Iliev}, {Mellema}, {Ahn}, {Shapiro}, {Mao}
  \& {Pen}}{{Iliev} et~al.}{2014}]{iliev2014}
{Iliev} I.~T.,  {Mellema} G.,  {Ahn} K.,  {Shapiro} P.~R.,  {Mao} Y.,   {Pen}
  U.-L.,  2014, \mn@doi [\mnras] {10.1093/mnras/stt2497}, \href
  {http://adsabs.harvard.edu/abs/2014MNRAS.439..725I} {439, 725}

\bibitem[\protect\citeauthoryear{{Kakiichi} et~al.,}{{Kakiichi}
  et~al.}{2018}]{kakiichi2018}
{Kakiichi} K.,  et~al., 2018, \mn@doi [\mnras] {10.1093/mnras/sty1318}, \href
  {http://adsabs.harvard.edu/abs/2018MNRAS.479...43K} {479, 43}

\bibitem[\protect\citeauthoryear{{Keating}, {Puchwein}  \&
  {Haehnelt}}{{Keating} et~al.}{2018}]{keating2018}
{Keating} L.~C.,  {Puchwein} E.,   {Haehnelt} M.~G.,  2018, \mn@doi [\mnras]
  {10.1093/mnras/sty968}, \href
  {https://ui.adsabs.harvard.edu/abs/2018MNRAS.477.5501K} {477, 5501}

\bibitem[\protect\citeauthoryear{{Konno} et~al.,}{{Konno}
  et~al.}{2018}]{konno2018}
{Konno} A.,  et~al., 2018, \mn@doi [\pasj] {10.1093/pasj/psx131}, \href
  {http://adsabs.harvard.edu/abs/2018PASJ...70S..16K} {70, S16}

\bibitem[\protect\citeauthoryear{{Kulkarni}, {Keating}, {Haehnelt}, {Bosman},
  {Puchwein}, {Chardin}  \& {Aubert}}{{Kulkarni} et~al.}{2019a}]{kulkarni2019}
{Kulkarni} G.,  {Keating} L.~C.,  {Haehnelt} M.~G.,  {Bosman} S.~E.~I.,
  {Puchwein} E.,  {Chardin} J.,   {Aubert} D.,  2019a, \mn@doi [\mnras]
  {10.1093/mnrasl/slz025}, \href
  {http://adsabs.harvard.edu/abs/2019MNRAS.485L..24K} {485, L24}

\bibitem[\protect\citeauthoryear{{Kulkarni}, {Worseck}  \&
  {Hennawi}}{{Kulkarni} et~al.}{2019b}]{kulkarni2019agn}
{Kulkarni} G.,  {Worseck} G.,   {Hennawi} J.~F.,  2019b, \mn@doi [Monthly
  Notices of the Royal Astronomical Society] {10.1093/mnras/stz1493}, \href
  {https://ui.adsabs.harvard.edu/abs/2019MNRAS.488.1035K} {488, 1035}

\bibitem[\protect\citeauthoryear{{Levermore}}{{Levermore}}{1984}]{levermore1984}
{Levermore} C.~D.,  1984, \mn@doi [\jqsrt] {10.1016/0022-4073(84)90112-2},
  \href {http://adsabs.harvard.edu/abs/1984JQSRT..31..149L} {31, 149}

\bibitem[\protect\citeauthoryear{{Lidz} \& {Malloy}}{{Lidz} \&
  {Malloy}}{2014}]{lidz2014}
{Lidz} A.,  {Malloy} M.,  2014, \mn@doi [\apj] {10.1088/0004-637X/788/2/175},
  \href {http://adsabs.harvard.edu/abs/2014ApJ...788..175L} {788, 175}

\bibitem[\protect\citeauthoryear{{Lidz}, {Oh}  \& {Furlanetto}}{{Lidz}
  et~al.}{2006}]{lidz2006}
{Lidz} A.,  {Oh} S.~P.,   {Furlanetto} S.~R.,  2006, \mn@doi [\apjl]
  {10.1086/502678}, \href {http://adsabs.harvard.edu/abs/2006ApJ...639L..47L}
  {639, L47}

\bibitem[\protect\citeauthoryear{{Lidz}, {McQuinn}, {Zaldarriaga}, {Hernquist}
  \& {Dutta}}{{Lidz} et~al.}{2007}]{lidz2007}
{Lidz} A.,  {McQuinn} M.,  {Zaldarriaga} M.,  {Hernquist} L.,   {Dutta} S.,
  2007, \mn@doi [\apj] {10.1086/521974}, \href
  {http://adsabs.harvard.edu/abs/2007ApJ...670...39L} {670, 39}

\bibitem[\protect\citeauthoryear{{Malloy} \& {Lidz}}{{Malloy} \&
  {Lidz}}{2015}]{lidz2015}
{Malloy} M.,  {Lidz} A.,  2015, \mn@doi [\apj] {10.1088/0004-637X/799/2/179},
  \href {http://adsabs.harvard.edu/abs/2015ApJ...799..179M} {799, 179}

\bibitem[\protect\citeauthoryear{{Mason}, {Treu}, {Dijkstra}, {Mesinger},
  {Trenti}, {Pentericci}, {de Barros}  \& {Vanzella}}{{Mason}
  et~al.}{2018}]{mason2018}
{Mason} C.~A.,  {Treu} T.,  {Dijkstra} M.,  {Mesinger} A.,  {Trenti} M.,
  {Pentericci} L.,  {de Barros} S.,   {Vanzella} E.,  2018, \mn@doi [\apj]
  {10.3847/1538-4357/aab0a7}, \href
  {http://adsabs.harvard.edu/abs/2018ApJ...856....2M} {856, 2}

\bibitem[\protect\citeauthoryear{{Mason} et~al.,}{{Mason}
  et~al.}{2019}]{mason2019}
{Mason} C.~A.,  et~al., 2019, \mn@doi [\mnras] {10.1093/mnras/stz632}, \href
  {http://adsabs.harvard.edu/abs/2019MNRAS.tmp..618M} {}

\bibitem[\protect\citeauthoryear{{McGreer}, {Mesinger}  \&
  {D'Odorico}}{{McGreer} et~al.}{2015}]{mcgreer2015}
{McGreer} I.~D.,  {Mesinger} A.,   {D'Odorico} V.,  2015, \mn@doi [\mnras]
  {10.1093/mnras/stu2449}, \href
  {http://adsabs.harvard.edu/abs/2015MNRAS.447..499M} {447, 499}

\bibitem[\protect\citeauthoryear{{Mesinger}}{{Mesinger}}{2010}]{mesinger2010}
{Mesinger} A.,  2010, \mn@doi [\mnras] {10.1111/j.1365-2966.2010.16995.x},
  \href {http://adsabs.harvard.edu/abs/2010MNRAS.407.1328M} {407, 1328}

\bibitem[\protect\citeauthoryear{{Mesinger}, {Aykutalp}, {Vanzella},
  {Pentericci}, {Ferrara}  \& {Dijkstra}}{{Mesinger}
  et~al.}{2015}]{mesinger2015}
{Mesinger} A.,  {Aykutalp} A.,  {Vanzella} E.,  {Pentericci} L.,  {Ferrara} A.,
    {Dijkstra} M.,  2015, \mn@doi [\mnras] {10.1093/mnras/stu2089}, \href
  {http://adsabs.harvard.edu/abs/2015MNRAS.446..566M} {446, 566}

\bibitem[\protect\citeauthoryear{{O{\~n}orbe}, {Davies}, {Luki{\'c}}, {},
  {Hennawi}  \& {Sorini}}{{O{\~n}orbe} et~al.}{2019}]{onorbe2019}
{O{\~n}orbe} J.,  {Davies} F.~B.,  {Luki{\'c}} {} Z.,  {Hennawi} J.~F.,
  {Sorini} D.,  2019, \mn@doi [\mnras] {10.1093/mnras/stz984}, \href
  {https://ui.adsabs.harvard.edu/abs/2019MNRAS.486.4075O} {486, 4075}

\bibitem[\protect\citeauthoryear{{Osterbrock} \& {Ferland}}{{Osterbrock} \&
  {Ferland}}{2006}]{osterbrock2006}
{Osterbrock} D.~E.,  {Ferland} G.~J.,  2006, {Astrophysics of gaseous nebulae
  and active galactic nuclei}.
University Science Books

\bibitem[\protect\citeauthoryear{{Pawlik} \& {Schaye}}{{Pawlik} \&
  {Schaye}}{2011}]{pawlik2011}
{Pawlik} A.~H.,  {Schaye} J.,  2011, \mn@doi [\mnras]
  {10.1111/j.1365-2966.2010.18032.x}, \href
  {http://adsabs.harvard.edu/abs/2011MNRAS.412.1943P} {412, 1943}

\bibitem[\protect\citeauthoryear{{Planck Collaboration}}{{Planck
  Collaboration}}{2018}]{planck2018}
{Planck Collaboration} 2018, arXiv:1807.06209, \href
  {http://adsabs.harvard.edu/abs/2018arXiv180706209P} {}

\bibitem[\protect\citeauthoryear{{Planck Collaboration XVI.}}{{Planck
  Collaboration XVI.}}{2014}]{planck2014}
{Planck Collaboration XVI.} 2014, \mn@doi [\aap] {10.1051/0004-6361/201321591},
  \href {http://adsabs.harvard.edu/abs/2014A%26A...571A..16P} {571, A16}

\bibitem[\protect\citeauthoryear{{Puchwein}, {Haardt}, {Haehnelt}  \&
  {Madau}}{{Puchwein} et~al.}{2019}]{puchwein2019}
{Puchwein} E.,  {Haardt} F.,  {Haehnelt} M.~G.,   {Madau} P.,  2019, \mn@doi
  [\mnras] {10.1093/mnras/stz222}, \href
  {http://adsabs.harvard.edu/abs/2019MNRAS.485...47P} {485, 47}

\bibitem[\protect\citeauthoryear{{Sadoun}, {Zheng}  \&
  {Miralda-Escud{\'e}}}{{Sadoun} et~al.}{2017}]{sadoun2017}
{Sadoun} R.,  {Zheng} Z.,   {Miralda-Escud{\'e}} J.,  2017, \mn@doi [\apj]
  {10.3847/1538-4357/aa683b}, \href
  {https://ui.adsabs.harvard.edu/abs/2017ApJ...839...44S} {839, 44}

\bibitem[\protect\citeauthoryear{{Shibuya} et~al.,}{{Shibuya}
  et~al.}{2018}]{shibuya2018}
{Shibuya} T.,  et~al., 2018, \mn@doi [\pasj] {10.1093/pasj/psx122}, \href
  {http://adsabs.harvard.edu/abs/2018PASJ...70S..14S} {70, S14}

\bibitem[\protect\citeauthoryear{{Springel}}{{Springel}}{2005}]{springel2005gadget}
{Springel} V.,  2005, \mn@doi [\mnras] {10.1111/j.1365-2966.2005.09655.x},
  \href {http://adsabs.harvard.edu/abs/2005MNRAS.364.1105S} {364, 1105}

\bibitem[\protect\citeauthoryear{{Steidel}, {Bogosavljevi{\'c}}, {Shapley},
  {Reddy}, {Rudie}, {Pettini}, {Trainor}  \& {Strom}}{{Steidel}
  et~al.}{2018}]{steidel2018}
{Steidel} C.~C.,  {Bogosavljevi{\'c}} M.,  {Shapley} A.~E.,  {Reddy} N.~A.,
  {Rudie} G.~C.,  {Pettini} M.,  {Trainor} R.~F.,   {Strom} A.~L.,  2018,
  \mn@doi [\apj] {10.3847/1538-4357/aaed28}, \href
  {http://adsabs.harvard.edu/abs/2018ApJ...869..123S} {869, 123}

\bibitem[\protect\citeauthoryear{{Tepper-Garc{\'{\i}}a}}{{Tepper-Garc{\'{\i}}a}}{2006}]{teppergarcia2006}
{Tepper-Garc{\'{\i}}a} T.,  2006, \mn@doi [\mnras]
  {10.1111/j.1365-2966.2006.10450.x}, \href
  {http://adsabs.harvard.edu/abs/2006MNRAS.369.2025T} {369, 2025}

\bibitem[\protect\citeauthoryear{{Trenti}, {Stiavelli}, {Bouwens}, {Oesch},
  {Shull}, {Illingworth}, {Bradley}  \& {Carollo}}{{Trenti}
  et~al.}{2010}]{trenti2010}
{Trenti} M.,  {Stiavelli} M.,  {Bouwens} R.~J.,  {Oesch} P.,  {Shull} J.~M.,
  {Illingworth} G.~D.,  {Bradley} L.~D.,   {Carollo} C.~M.,  2010, \mn@doi
  [\apjl] {10.1088/2041-8205/714/2/L202}, \href
  {http://adsabs.harvard.edu/abs/2010ApJ...714L.202T} {714, L202}

\bibitem[\protect\citeauthoryear{{Viel}, {Haehnelt}  \& {Springel}}{{Viel}
  et~al.}{2004}]{viel2004}
{Viel} M.,  {Haehnelt} M.~G.,   {Springel} V.,  2004, \mn@doi [\mnras]
  {10.1111/j.1365-2966.2004.08224.x}, \href
  {http://adsabs.harvard.edu/abs/2004MNRAS.354..684V} {354, 684}

\bibitem[\protect\citeauthoryear{{Walther}, {O{\~n}orbe}, {Hennawi}  \&
  {Luki{\'c}}}{{Walther} et~al.}{2019}]{walther2019}
{Walther} M.,  {O{\~n}orbe} J.,  {Hennawi} J.~F.,   {Luki{\'c}} Z.,  2019,
  \mn@doi [\apj] {10.3847/1538-4357/aafad1}, \href
  {http://adsabs.harvard.edu/abs/2019ApJ...872...13W} {872, 13}

\bibitem[\protect\citeauthoryear{{Weinberger}, {Haehnelt}  \&
  {Kulkarni}}{{Weinberger} et~al.}{2019}]{weinberger2019}
{Weinberger} L.~H.,  {Haehnelt} M.~G.,   {Kulkarni} G.,  2019, \mn@doi [\mnras]
  {10.1093/mnras/stz481}, \href
  {http://adsabs.harvard.edu/abs/2019MNRAS.485.1350W} {485, 1350}

\bibitem[\protect\citeauthoryear{{Worseck} et~al.,}{{Worseck}
  et~al.}{2014}]{worseck2014}
{Worseck} G.,  et~al., 2014, \mn@doi [\mnras] {10.1093/mnras/stu1827}, \href
  {http://adsabs.harvard.edu/abs/2014MNRAS.445.1745W} {445, 1745}

\makeatother
\end{thebibliography}

\bsp

\label{lastpage}

\end{document}